# Extraction of higher-order nonlinear electronic response to strong field excitation in solids using high harmonic generation

Seunghwoi Han[1]**, Lisa Ortmann[2], Hyunwoong Kim[1], Yong Woo Kim[1], Takashi Oka[2,3], Alexis Chacon[4], Brent Doran[5], Marcelo Ciappina[6], Maciej Lewenstein[7,8], Seung-Woo Kim[1]*, Seungchul Kim[9]*, Alexandra S. Landsman[2]

L.O. , S.H. and H.K. contributed equally to this work.

[1] Department of Mechanical Engineering, Korea Advanced Institute of Science and Technology (KAIST), 291 Daehak-ro, Yuseong-gu, Daejeon 34141, South Korea

[2] Max Planck Institute for the Physics of Complex Systems, Noethnitzer Strasse 38, 01187 Dresden, Germany

[3] Max Planck Institute for Chemical Physics of Solids, Noethnitzer Strasse 40, 01187 Dresden, Germany

[4] Center for Nonlinear Studies, Physics and Chemistry of Materials T1-Division, Los Alamos National Laboratory, Los Alamos NM 87545, USA

[5] Mathematical Institute, University of Oxford, Oxford, United Kingdom

[6] Institute of Physics of the ASCR, ELI-Beamlines, Na Slovance 2, 182 21 Prague, Czechia

[7] ICFO – Institut de Ciencies Fotoniques, The Barcelona Institute of Science and Technology, Av. Carl Friedrich Gauss 3, 08860 Castelldefels (Barcelona), Spain

[8] ICREA, Pg. Lluis Company 23, 08010 Barcelona, Spain

[9] Department of Optics and Mechatronics Engineering, College of Nanoscience and Nanotechnology, Pusan National University, 2 Busandaehak-ro 63beon-gil, Busan 46241, South Korea

*Corresponding authors: Seungchul Kim (s.kim@pusan.ac.kr) and Seung-Woo Kim (swk@kaist.ac.kr)

**Current position: Institute for the Frontier of Attosecond Science and Technology, CREOL and Department of Physics, University of Central Florida, Orlando, Florida 32816, USA

**Abstract**

State-of-the-art experiments employ strong ultrafast optical fields to study the nonlinear response of electrons in solids on an attosecond time-scale[1–13]. Notably, a recent experiment[1] retrieved a 3rd order nonlinear susceptibility by comparing the nonlinear response induced by a strong laser field to a linear response induced by the otherwise identical weak field. In parallel, experiments have demonstrated high harmonic generation (HHG) in solids[2-11], a highly nonlinear process that until recently had only been observed in gases. The highly nonlinear nature of HHG has the potential to extract even higher order nonlinear susceptibility terms, and thereby characterize the entire response of the electronic system to strong field excitation. However, up till now, such characterization has been elusive due to a lack of direct correspondence between high harmonics and nonlinear susceptibilities. Here, we demonstrate a regime where such correspondence can be clearly made, extracting nonlinear susceptibilities (7th, 9th, and 11th) from sapphire of the same order as the measured high harmonics. The extracted high order susceptibilities show angular-resolved periodicities arising from variation in the band structure with crystal orientation. Nonlinear susceptibilities are key to ultrafast lightwave driven optoelectronics, allowing petahertz scaling manipulation of the signal[1,14]. Our results open a door to multi-channel signal processing, controlled by laser polarization.

The development of strong ultrafast optical fields opened a possibility to induce and observe the nonlinear dynamics of electrons without damaging the material[12,13]. High harmonic generation in solids is a nonlinear frequency conversion phenomenon, whose mechanism is still a subject of intense investigation[2-11,15]. The process was first observed in atomic gas[16], where it is well-described by the three-step recollision model[17–19]. However, in solids, in addition to the above-mentioned recollision scenario (described within non-perturbative interband HHG framework), intraband transitions[3,4] and Wannier-Stark localization[10] have been found to make important contributions, depending on the material and laser parameters.

Prior experimental extraction of nonlinear susceptibility in attosecond experiments relied on other, non-HHG, mechanisms, such as comparing the linear to the nonlinear polarization response using attosecond streak camera[1]. This allowed the extraction of the third order Kerr nonlinearity, showing the potential of petahertz scaling manipulation by relying on the change of the effective refractive index in optoelectronics[1,20]. However, extracting higher order nonlinear susceptibilities using this method requires broader spectral coverage. Here we show that in fact it is possible to use HHG in suitable regimes to cleanly extract much higher order susceptibilities and therefore achieve unprecedented characterization of nonlinear electronic response directly based on experimental measurements.

Figure 1 shows different possible mechanisms of HHG in solids, which may be due to either interband transitions (resulting from multi-photon absorption or tunneling) or intraband transitions. Both interband and intraband HHG have been previously observed in solids[2-7], with recollision being directly analogous to HHG in atomic gas[21], while intraband oscillations being unique to solids and determined by the shape of the conduction band[22]. Here, we demonstrate high harmonics produced by the multi-photon interband transitions. Furthermore, we show the existence of (i) multi-photon scaling and (ii) interband polarization leading to high harmonics creates the necessary and sufficient conditions for measuring high order non-linear susceptibilities, which are inaccessible by other means.

In our experiment, a laser source emitting 12-fs, infrared laser pulses (800 nm central wavelength) at a 75 MHz repetition rate is used. The ultrashort pulses are focused on an effective spot size of ~ 2 μm, with peak intensities up to 12 TW/cm$^2$. The incident pulse power is varied using waveplates in combination with polarizers. The polarization direction of the incident pulses is fixed to be linear. The samples are mounted on a 3D translational stage equipped with an extra rotation axis to vary the crystallographic angle with respect to the incident linear polarization. The samples are made of single crystal sapphire wafers with a thickness of 430 μm, being cut along the C- and A-plane. Figure 2 shows the 7th to 13th order harmonics over a range of rotational angles with increasing laser field strength. As a photomultiplier was used which responds to EUV radiation with wavelength range of 45-135 nm, which corresponds to harmonic orders from 7 to 17, the harmonic spectrum from order 1 to 5, which corresponds to below band-gap harmonics, is not depicted. Figure 2 shows the measured high harmonic yield depends on the crystallographic orientation, which was analyzed by rotating the specimen about the polarization direction of the incident laser field. There are three distinct directions, for which the laser is polarized along the Γ-K, Γ-M or Γ-A direction, and which we will focus on in part of the theoretical analysis due to the easy accessibility of the band structure in these directions.

The multi-photon regime is characterized by the relatively high value of the Keldysh parameter, $\gamma = \omega \frac{\sqrt{2m\Delta}}{eF}$, where $\omega, F, \Delta$ are the laser frequency, maximum field strength and bandgap, respectively, e > 0 is the absolute value of the electron charge, and m is the reduced mass of an electron and a hole ($m^{-1} = m_e^{-1} + m_h^{-1}$).

The high bandgap of sapphire (8.8 eV), combined with relatively high frequency and low peak field strength leads to a high value of the Keldysh parameter corresponding to $\gamma \sim 2.6$ (depending on the exact intensity), which is consistent with the multi-photon regime. In this regime, the measured harmonics follow perturbative scaling, given by $I_{HHG} \propto I^N$, where $I_{HHG}$ is the intensity of the measured harmonics and $N$ is their order. This perturbative scaling is observed for all crystal orientations, as Figure 2c shows. The flattening out of the curve at higher intensities observed in Fig. 2c is consistent with the transition from the multi-photon to tunneling regime. The remaining deviation observed for the 7th harmonic is due to quantum path interference, explained in some detail in [23]. This quantum path interference also influences the 9th and 11th order harmonic yields, but the effect is less pronounced, since the distortion of the generated harmonics is different for each N order and depends on intensity.[23] The power-law scaling with intensity of the laser is well-known for multi-photon processes, unambiguously confirming that the experiment takes place in the multi-photon regime.

Figure 2(a) depicts the measured spectra by rotating the laser polarization about the C-plane. The electric field induced inside the sapphire specimen reaches to 0.7 V/Å. The harmonic yield has a six-fold symmetry (periodicity of 60 degrees), which is same as the periodicity of crystal orientation in C-plane sapphire. The 7th, 11th, and 13th order harmonics have the maximum yield in the Γ-K direction, and only the 9th order peak has the maximum value in the Γ-M direction. These differences reflect varying high order nonlinear susceptibilities for different crystal orientations. Figure 2(b) shows the high harmonic yield of A-plane sapphire as a function of crystal orientation, which varies from Γ-A to Γ-M directions. The measured harmonics have maximum yields in the Γ-M direction and local maximum yields in the Γ-A direction, showing a two-fold symmetry.

The ordinary and extraordinary refractive indices of sapphire are known to be 1.7601 and 1.7522, respectively, for 800 nm in wavelength. The birefringence effects become important for A-plane thick sapphire when the polarization direction is located between G-M (ordinary) and G-A (extraordinary) directions. However, for C-plane sapphire, the light propagates along the optic axis, where the refractive index is rotationally invariant, allowing one to neglect the birefringence effects. Additionally, the residual polarization status was compensated with the combination of wave plates.

The sapphire crystal is a dense periodic bulk solid and high harmonic generation mainly takes place along the micro scale focused depth of the pump laser. However, sapphire can act as both an HHG emitter and a strong EUV absorber, so that the observed HHG is produced within a few tens of nanometers. The small volume involved implies that phase matching here is less critical than in a gas target.

HHG in solids is commonly interpreted as being due to either inter- or intraband contributions[2-9]. In panels (a) and (c) of Figure 3 the calculated inter- and intraband contributions to the high harmonic yield are depicted for the laser polarization pointing along the Γ-K direction. As the comparison of panels (a) and (c) in Figure 3 reveals, interband harmonics dominate over intraband above the bandgap. The implementation included a total of 5 bands (see SI for details), which were chosen based on the dominant transition dipoles. The band structure calculations were performed using WIEN2k program package[24], with the bandgap taken to be 8.8 eV, in agreement with known values.

The most compelling evidence against the intraband mechanism, independent of fine details of any computational approaches, comes from comparing the experimentally measured high harmonic yield along the Γ-A direction against well-known qualitative expectations for intraband harmonics[2,3,22]. In particular, as Figure 3b shows, the measured harmonic intensity is comparable in all three crystal directions. On the other hand, intraband currents would contribute much higher frequency components along the Γ-A direction, relative to the Γ-K and Γ-M directions (see Fig. 3b). Physically, this is due to the k-vector being significantly shorter along the direction perpendicular to the hexagonal face of the lattice (see Fig. 2 for crystal structure and Fig. 2d for band structure of sapphire).

The presence of higher frequency components along Γ-A (relative to other directions) becomes clear if one expands the conduction band as a sum $E_i(k) = \sum_{n=0}^{6} \epsilon_{n,i} \cos(nka_i)$ with fitting constants $\epsilon_n$ and lattice constants $a_i$ [22,25]. The cosine sum takes account of intraband HHG emission, which is due to non-parabolic nature of the dispersion curve in the conduction band[2,22]. Since the dispersion curve along the Γ-A direction is composed of much higher frequency components, intraband harmonics along the Γ-A direction have a significantly higher cut-off than along other directions, as is indeed shown in Fig. 3b. The absence of this in the measured yield (where all directions have the same cut-offs around the 13th harmonic) indicates that intraband oscillations are not the dominant mechanism producing high harmonics in this experiment. In contrast, the interband model predicts all crystal directions to have similar cut-offs, in agreement with experimental observations (see Fig. 3d).

Having established that the measured harmonics are due to interband transitions in the multi-photon regime, we can now compute high-order nonlinear susceptibilities. The formula for polarization along the laser direction, neglecting tensorial and non-instantaneous effects, is given by:

$$p(t) = \epsilon_0 V_{focal} \sum_i \chi^{(i)} \cdot E^i(t) \quad (1)$$

with $\epsilon_0$ denoting the vacuum permittivity and $\chi^{(N)}$ the susceptibility of order $N$ and where we assumed that the polarization density is constant over the focal volume $V_{focal}$. Here, we use the expansion

$$E^i(t) = E_0^i \cos^i(\omega_0 t) = E_0^i \sum_{j=1}^{i} a_j \cos(j\omega_0 t) \quad (2)$$

The above electric field is valid within the dipole approximation, which assumes the spatial variation of the laser field is small relative to the spatial excursion of the electron. This is indeed the case here, where the excursion of the electron is on the order of nanometers (note that in the sapphire crystal, the conventional unit cell has the size of ~1 nm), compared to the 800 nanometer laser wavelength.

The harmonic yield due to interband transitions is given by the Fourier transform of the polarization [15]

$$Yield = \left| FT\left\{ \frac{d^2}{dt^2} p(t) \right\} \right|^2 \quad (3)$$

where the Nth harmonic is given by the Nth Fourier component. Since the experimentally measured harmonics satisfy power-law scaling (see Fig. 2c), given by

$$Yield_N = A_N I^N \propto A_N \cdot E_0^{2N} \quad (4)$$

the Nth order yield is calculated plugging in

$$p_N(t) \propto \epsilon_0 E_0^N \chi^{(N)} \cos(N\omega_0 t) \quad (5)$$

into eq. 3 obtaining

$$Yield_N \propto \epsilon_0^2 |\chi^{(N)}|^2 E_0^{2N} \propto \epsilon_0^2 |\chi^{(N)}|^2 I^N \quad (6)$$

Consequently, $\chi^{(N)}$ is contained in the prefactor $A_N$ (see eq. 4) and thus, taking care of all the further constants and prefactors (see SM for details) the experimentally determined $A_N$ value can be used to obtain nonlinear susceptibilities $\chi^{(N)}$ directly from experimental measurements. Note that the above scheme for calculating nonlinear susceptibilities requires both interband mechanism for high harmonics (to satisfy eq. 3) and power-law scaling (characteristic of multi-photon transitions) that satisfies eq. 4. Therefore, it could not be applied to many prior experiments, where HHG was due to either intraband, or tunneling interband transitions.

In Figure 4, the nonlinear susceptibilities extracted using experimentally measured yields for the 7th, 9th and 11th order harmonics in the C-plane are shown as a function of crystal orientation. The shape of the harmonics is affected by the broad bandwidth of the 12 fs driving pulse, which in turn, leads to "effective" susceptibilities that reflect the mixing of the whole bandwidth. The fact that the values are approximately the same for 0 and 60 degrees is due to the 6-fold symmetry of the crystal along the C-plane. The absolute values of the nonlinear susceptibilities are found to be on the order of $10^{-67}$ for the 7th harmonic, $10^{-87}$ for the 9th harmonic and $10^{-107}$ for the 11th harmonic, leading to a constant ratio between consecutive order susceptibilities, $\chi^{(N)}/\chi^{(N+2)}$, on the order of $10^{-20}$. Both the absolute and relative values are in line with the approximate ball-park values found in the literature, where the Nth order nonlinear susceptibility of solids is estimated to be on the order of $(5 \cdot 10^{11} \text{ V/m})^{-N+1}$ [14]. Note that measured susceptibilities decline less rapidly with harmonic order than this generic ballpark estimate, suggesting nonlinear contributions may play a greater role in sapphire than some other materials.

In conclusion, we investigated nonlinear electronic response in wide bandgap material by measuring angular-dependent high harmonic emission. Having established that the high harmonics are due to interband transitions in the multi-photon regime, we were able to extract orientation-dependent high-order nonlinear susceptibilities of the material. This greatly expands on prior findings, which extracted the 3rd order Kerr susceptibility induced by strong optical fields[1]. Higher order nonlinearities are believed to be crucial to signal manipulation in optoelectronics by affecting the electron response time and refractive index[14,26]. The susceptibilities obtained in our study have a periodicity that depends on crystal orientation, suggesting a possibility of multichannel signal processing at PHz frequencies by controlling the electron response time and refractive index using laser polarization.

## Acknowledgements

ASL acknowledges the Max Planck Center for Attosecond Science (MPC-AS). A.C. performed part of the work under the auspices of the Los Alamos National Laboratory, which is operated by LANS, LLC, for the NNSA for the US DOE under contract No. DE-AC52-06NA25396. M. C. acknowledges the project Advanced research using high intensity laser produced photons and particles (CZ.02.1.01/0.0/0.0/16\_019/0000789) from European Regional Development Fund (ADONIS). This work was supported by the National Research Foundation of the Republic of Korea (NRF-2012R1A3A1050386), ICT, Future Planning (NRF-2017M3D1A1039287), and Basic Research Lab Program (NRF-2018R1A4A1025623)

## Author Contributions

The project was planned and overseen by S.K., S.-W.K. and A.S.L. High harmonic generation experiments were performed by S.H., H.K. and Y.W.K. Data was analyzed by S.H and Y.W.K. Simulations were performed by L.O. and A.C. All authors contributed to the discussion and preparation of the manuscript.

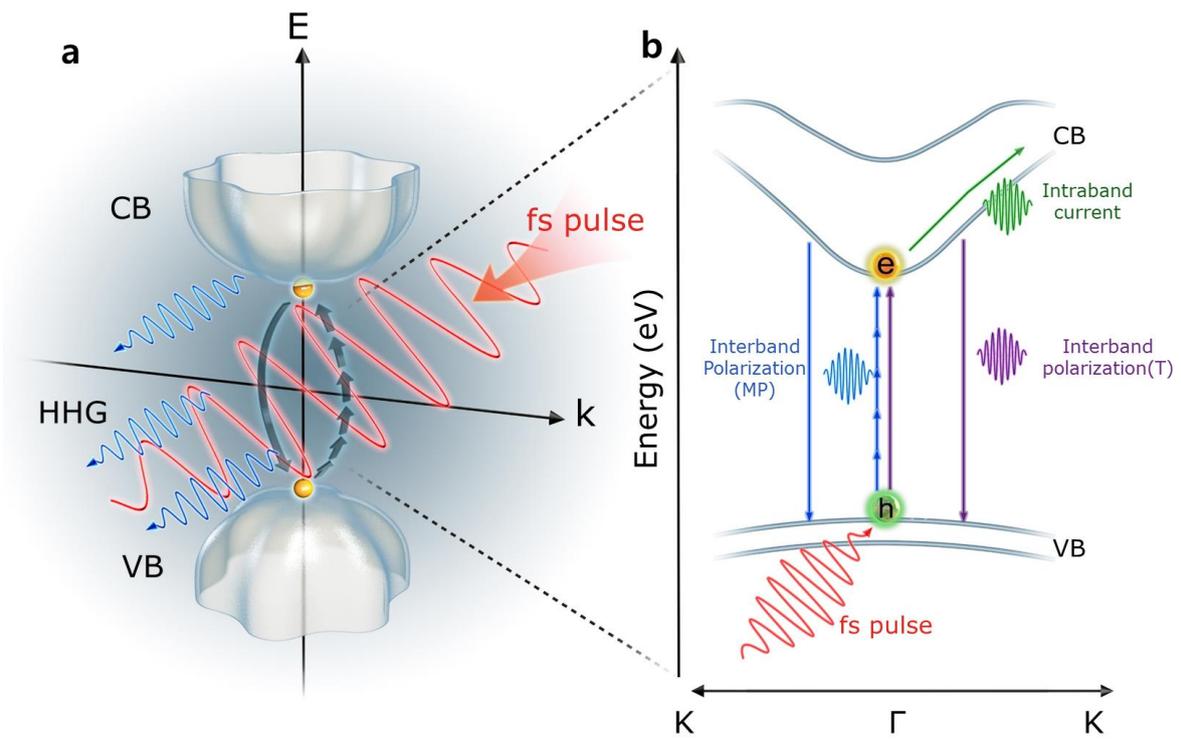

Figure 1. Schematic sketch of the interband and intraband mechanism. (a) Sketch of the 3D band structure with the valence band (VB) at the bottom and the conduction band (CB) at the top. The red wiggle represents the input laser field, and the blue wiggles the high harmonic radiation. (b) The 1D band structure with tunnel (T) transition and multiphoton (MP) transition as possible transition paths from the valence to the conduction band. Intraband HHG is due to the electron's oscillation in the conduction band, whereas interband HHG depends on the electron recombining with its hole.

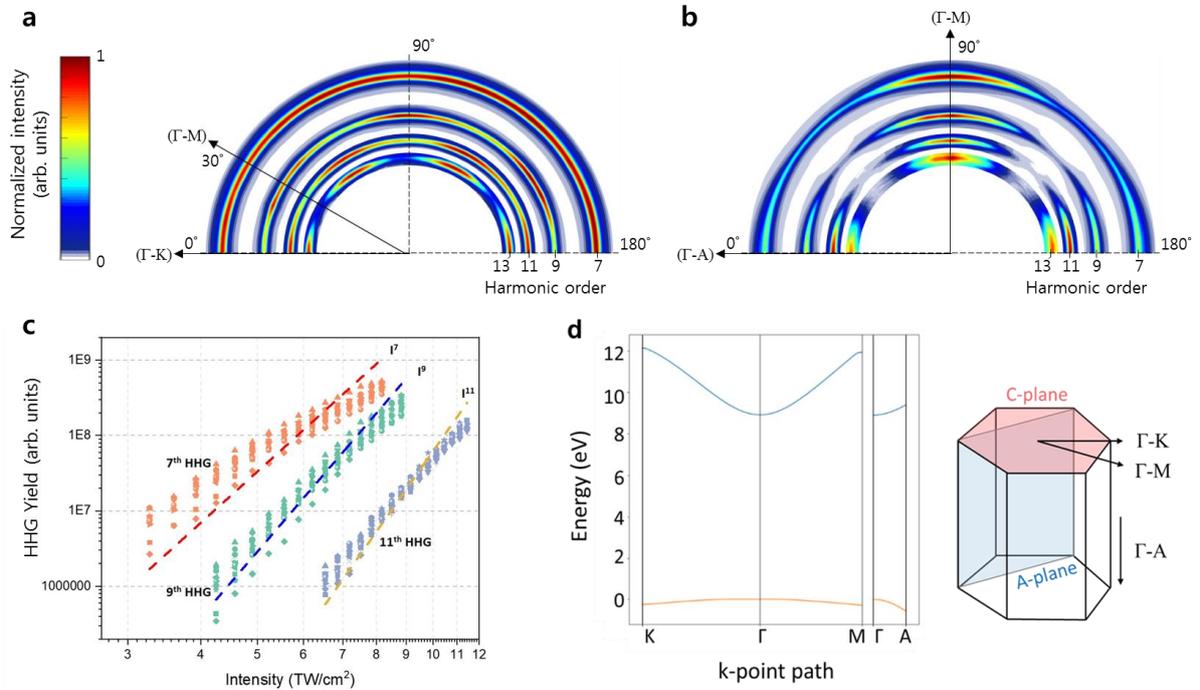

Figure 2. Experimental data: High harmonic generation yield sweeping over crystal orientation in C-plane (a) and A-plane (b) for various intensities. Each harmonic is normalized by its maximum intensity over crystal orientation so that the six-fold symmetry in C-plane and two-fold symmetry in A-plane sapphire are more visible. (a) Measured high harmonic spectra by rotating the laser polarization direction from Γ-K to Γ-M to Γ-K. The harmonic intensity in C-plane shows a six-fold symmetry. (b) Measured high harmonic spectra by rotating the laser polarization direction from Γ-A to Γ-M. The harmonic intensity in A-plane shows a two-fold symmetry. (c) The experimental data on a log-log scale: High harmonic generation yield sweeping over crystal orientation in C-plane for various intensities. The crystal orientation is rotated from 0 to 60 degree with the angle step of 5 degree.(12 data sets in total) The dashed lines represent fittings of the type $A \cdot I^N$ with $A$, the fitting parameter and $N$, the harmonic order as it is described in eq. (4). (red: 7th, blue: 9th, orange: 11th order) (d) The band structure of a sapphire crystal with the highest valence band and the lowest conduction band.

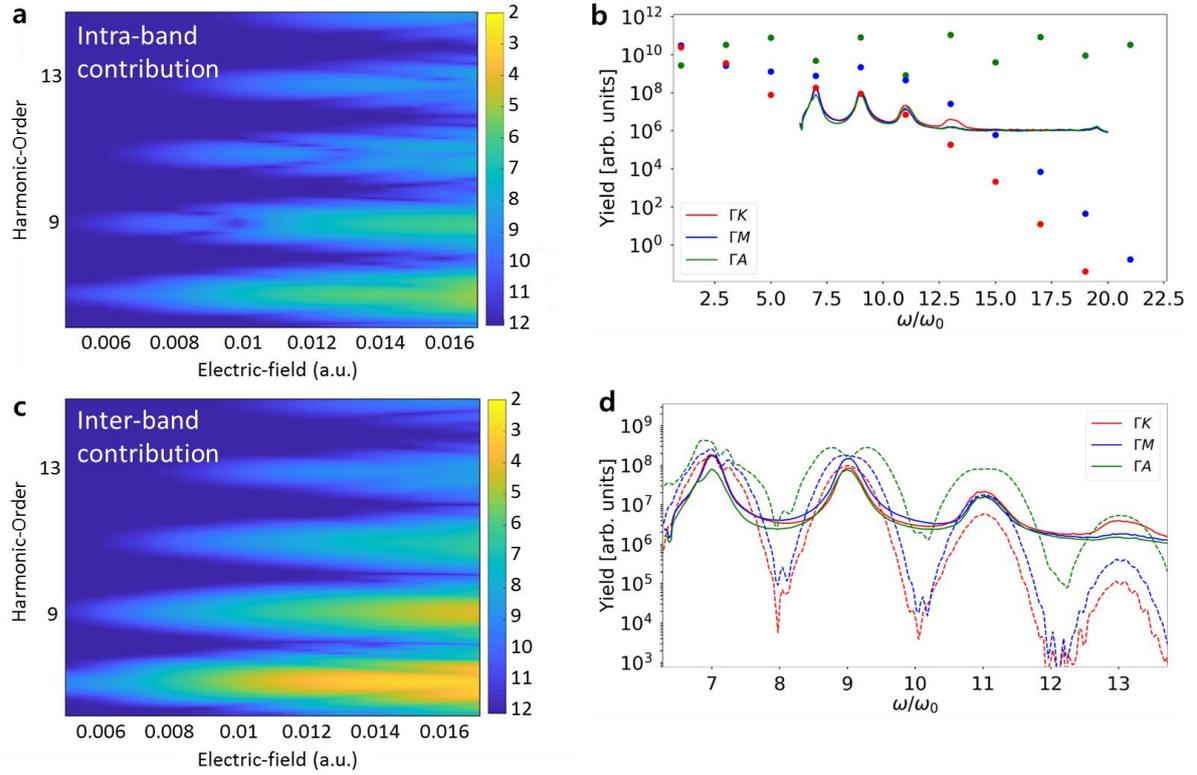

Figure 3. (a) and (b) show calculated intraband contributions, whereas (c) and (d) depict interband results. For direct comparison, the experimental data for different orientations is shown as solid lines in (b) and (d). Comparison of (a) and (c) shows that interband dominates over intraband contributions for all intensities along the ΓK direction. (b) The dots show harmonic intensity due to intraband transitions for three distinct crystal orientations (ΓA, ΓM, ΓK), calculated following [2]. While the other two directions have similar cut-offs, intraband transitions produce much higher harmonics along the ΓA direction, in contradiction to experimental results. This provides an independent confirmation that intraband contributions are unlikely to explain experimental measurements. (d) The dashed lines show interband contributions. Maximum harmonic orders are comparable for all three orientations, showing up to the 13$^{th}$ harmonic, in agreement with experiment. Calculations in (b) and (d) were done close to the experimentally calibrated field strength of $F_{0,exp}$=0.014 atomic units ($F_0$ = 0.0112 and 0.013 atomic units for intra and interband, respectively). The high harmonic yields in between (b) and (d) are not directly comparable because of different numerical approaches. For details on computational methods, see the supplement.

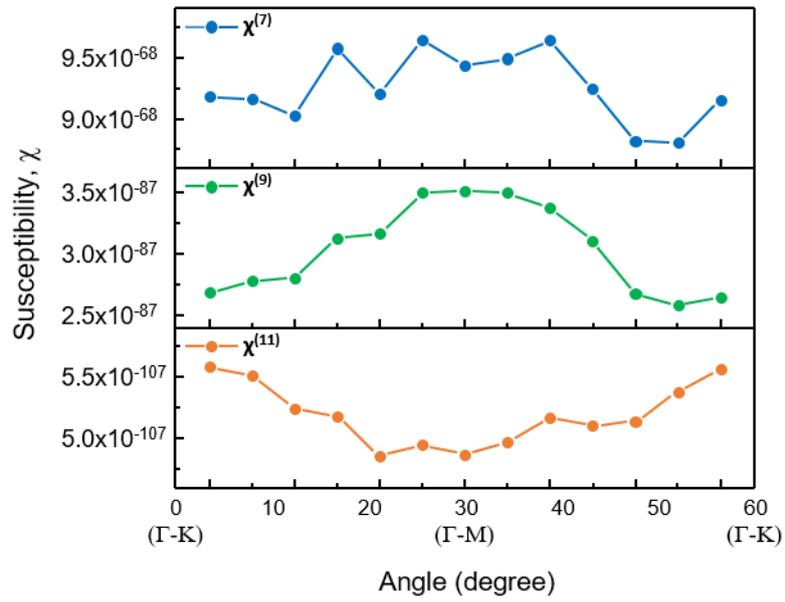

Figure 4. Angle-dependent high-order (7th, 9th and 11th) nonlinear susceptibilities calculated from the respective orders of the HHG yield.